\documentclass{article}
\usepackage{amsmath}
\usepackage{graphicx}
\usepackage{bm}
\begin{document}
%%%%%%%%%%%%%%%%%%%%%%%%%%%%%%%%%%%%%%%%%%%%%%%%%%%%%%%%%%%%%%%%%%%%%%
\title{
Analyzing collisions in classical mechanics 
using mass-momentum diagrams 
}
%%%%%%%%%%%%%%%%%%%%%%%%%%%%%%%%%%%%%%%%%%%%%%%%%%
\author{Akihiro Ogura}
\date{}
\maketitle
%%%%%%%%%%%%%%%%%%%%%%%%%%%%%%%%%%%%%%%%%%%%%%%%%%
\begin{center}
Laboratory of Physics, Nihon University, Matsudo 271-8587, Japan
\end{center}
%%%%%%%%%%%%%%%%%%%%%%%%%%%%%%%%%%%%%%%%%%%%%%%%%%

\vspace{2cm}

\begin{abstract}
%%%%%%%%%%%%%%%%%%%%%%%%%%%%%%%%%%%%%%%%%%%%%%%%%%%
We show the value of mass-momentum diagrams for analyzing collision 
problems in classical mechanics in one dimension. 
Collisions are characterized by the coefficient of restitution and 
the momentum of the interacting particles both before and after the 
collision. All those quantities are presented in the mass-momentum 
diagrams without the need to do any calculations. 
We also show that the same diagrams can be used to investigate 
collisions with respect to the center-of-mass system. In this case, 
also, we do not need to do any calculations to obtain the momentum. 
Since we give an alternative way of looking at collisions in 
classical mechanics, this article is aimed at undergraduate-level 
students. 

%%%%%%%%%%%%%%%%%%%%%%%%%%%%%%%%%%%%%%%%%%%%%%%%%%%
\end{abstract}

%%%%%%%%%%%%%%%%%%%%%%%%%%%%%%%%%%%%%%%%%%%%%%%%%%%%%%%%%%%%%%%%%%%%%%
\newpage
\section{Introduction}

Collisions between interacting particles are fundamentally important 
in physics, not only Newtonian mechanics but also relativistic 
mechanics. 
In many classroom problems concerning collisions, the masses, 
the velocities 
before collision and the coefficient of restitution of the colliding 
particles are given. The students are required to find the 
velocities of the particles after the collision. 
For relativistic mechanics, collisions are well 
described in an intuitive and instructive way using Minkowsky 
diagrams in momentum space~\cite{saletan, bokor, landau2}. 

In this article, we use this intuitive and instructive way for 
collisions in Newtonian mechanics. For this purpose, we introduce 
a mass-momentum diagram~\cite{takeuchi} with mass $m$ 
along the vertical axis and momentum $p$ represented on 
the horizontal axis. In this diagram, a point $(m, p)$ expresses 
the state of a particle. This diagram presents us with all the 
information we need about the collision without any calculations.

Moreover, the center-of-mass system is also used in collision 
problems. Using this makes equations simpler and 
calculations easier. Even in this case, 
we can obtain the momenta of the colliding particles from the 
mass-momentum diagrams without any calculations. 

This paper is organized in the following way. 
In Section 2, we recall one dimensional collisions with calculations. 
In Section 3, we introduce the mass-momentum 
diagrams and deduce all the equations from the diagrams without 
requiring any calculations. 
Next, we move to the center-of-mass system. We obtain all the 
quantities given in the center-of-mass system from the same 
mass-momentum diagram. 
In Section 4, we apply this diagram to the laboratory system. 
Section 5 is devoted to a summary.

%%%%%%%%%%%%%%%%%%%%%%%%%%%%%%%%%%%%%%%%%%%%%%%%%%%%%%%%%%%%%%%%%%%%%%
\section{Collisions between particles in one dimension 
with calculations}

Let us recall here the equations for one dimensional collisions. 
We consider that two particles, whose mass are $m_{A}$ and $m_{B}$, 
have velocities $v_{A}$ and $v_{B}$($v_{A}>v_{B}$) before the 
collision. We distinguish by primes the variables after 
the collisions. 

We write down the conservation of momentum and the definition of the 
coefficient of restitution $e$, 
\begin{equation}
\begin{cases}
&m_{A}v_{A}+m_{B}v_{B} = m_{A}v^{\prime}_{A}+m_{B}v^{\prime}_{B}, \\
&e = -\dfrac{v^{\prime}_{A}-v^{\prime}_{B}}{v_{A}-v_{B}}, 
\end{cases}
\label{eq:mocore}
\end{equation}
where the coefficient of restitution $e$ takes the value from 
$0$ to $1$. $e=0$ signifies a perfectly inelastic collision in which 
two particles are combined after collision, while $e=1$ a perfectly 
elastic collision in which the total energy is conserved. 
$0<e<1$ signifies a inelastic collision. 

Eqs.(\ref{eq:mocore}) give the velocities after collision 
\begin{equation}
\begin{cases}
v^{\prime}_{A} &= v_{A}-(1+e)\dfrac{m_{B}}{m_{A}+m_{B}}(v_{A}-v_{B}), \\
v^{\prime}_{B} &= v_{B}+(1+e)\dfrac{m_{B}}{m_{A}+m_{B}}(v_{A}-v_{B}),  
\end{cases}
\label{eq:avelo}
\end{equation}
and the momenta of the particles after collision 
\begin{equation}
\begin{cases}
p^{\prime}_{A} &= p_{A}-(1+e)\dfrac{m_{A}m_{B}}{m_{A}+m_{B}}(v_{A}-v_{B}), \\
p^{\prime}_{B} &= p_{B}+(1+e)\dfrac{m_{A}m_{B}}{m_{A}+m_{B}}(v_{A}-v_{B}), 
\end{cases}
\label{eq:amomentum}
\end{equation}
where the second term is the impulse experienced by each particles, 
which are equal and opposite, as fundamentally derived from 
Newton's third law. 
We clearly see the conservation of momentum $p_{A}+p_{B}=p'_{A}+p'_{B}$ 
from this expression for momentum. 
Note that $\frac{m_{A}m_{B}}{m_{A}+m_{B}}$ is a reduced mass and 
$v_{A}-v_{B}(>0)$ is a relative velocity before the collision.

%%%%%%%%%%%%%%%%%%%%%%%%%%%%%%%%%%%%%%%%
\section{Mass-momentum diagrams}

\begin{figure}[htbp]
\centering
\includegraphics[width=7.5cm]{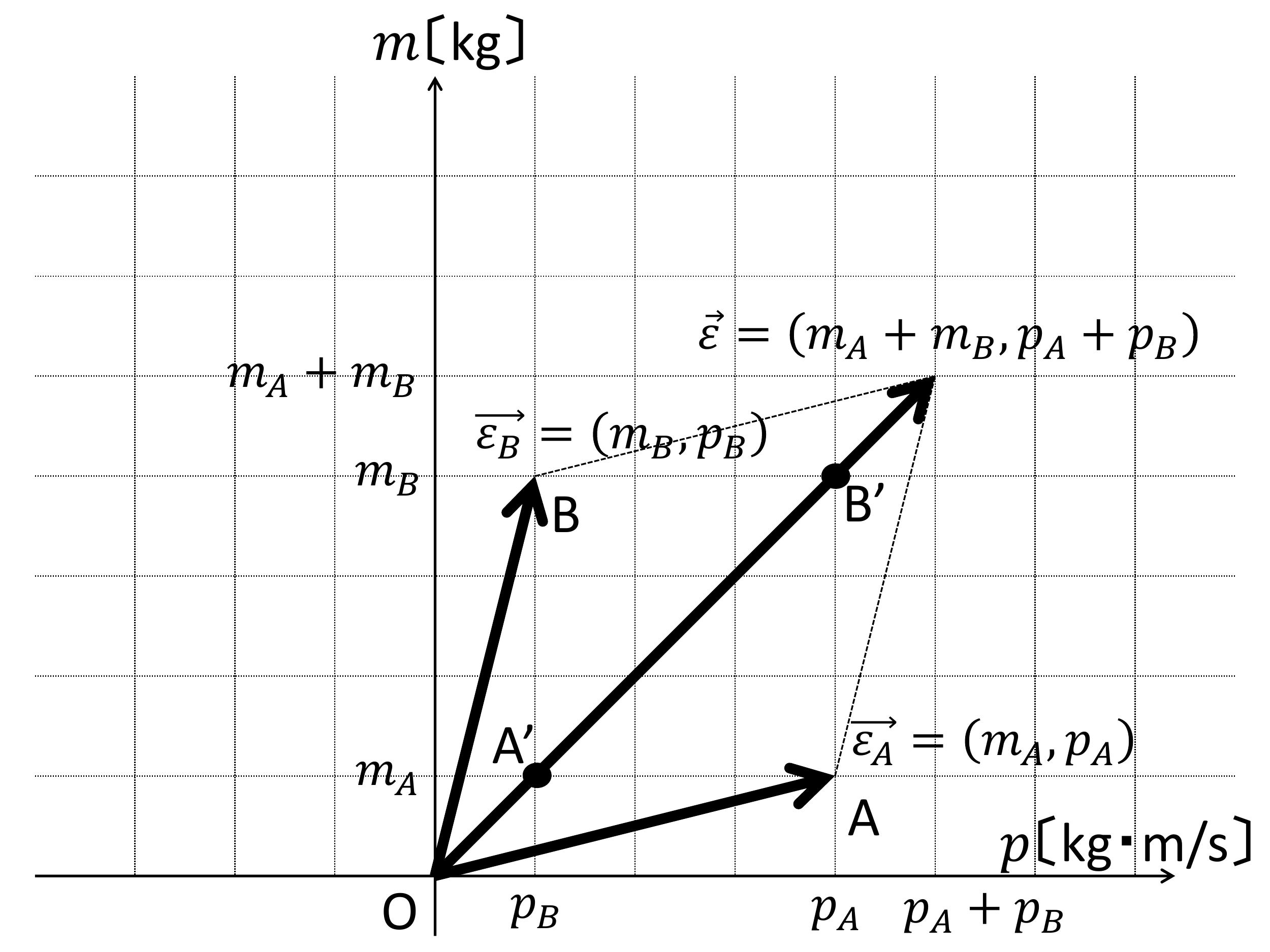}
\caption{The two vectors before collision. The case 
$(m_{A}, p_{A})=(1, 4)$ and $(m_{B}, p_{B})=(4, 1)$ means $v_{A}=4$ 
and $v_{B}=\frac{1}{4}$. In a perfectly inelastic collision 
($e = 0$), the two particles combine into a single particle which 
has mass $m_{A} + m_{B} = 5$ and momentum $p_{A} + p_{B} = 5$. 
Eq. (\ref{eq:lostgain}) shows AA${}^{\prime}$ = BB${}^{\prime}$ = 3. }
\label{fig:fig0203}
\end{figure}

In order to analyze the collisions in the previous section, 
we introduce the mass-momentum diagram which has 
the mass $m$ along the vertical axis and momentum $p$ 
represented by the horizontal axis~\cite{takeuchi}. 
If we extend this diagram to the special relativity, the vertical 
axis switches to energy instead of mass~\cite{saletan, bokor}. 
In this diagram, a point 
$(m, p)$ expresses the state of a particle. The interacting 
particles before a collision are represented by two 
vectors 
\begin{equation}
\begin{cases}
\vec{\varepsilon}_{A} &= (m_{A}, p_{A}), \\
\vec{\varepsilon}_{B} &= (m_{B}, p_{B}), 
\end{cases}
\end{equation}
as shown in figure \ref{fig:fig0203}. 

%%%%%%%%%%%%%%%%%%%%%%%%%%%%%%
%%%%%%%%%%%%%%%%%%%%%%%%%%%%%%

Now we consider the addition of these two vectors, 
\begin{equation}
\vec{\varepsilon} = \vec{\varepsilon}_{A}+\vec{\varepsilon}_{B} 
= (m_{A}+m_{B}, p_{A}+p_{B}), 
\end{equation}
which is depicted in figure \ref{fig:fig0203}. This is obtained using 
the parallelogram law. This vector $\vec{\varepsilon}$ has three 
characteristics which are described separately in following 
subsections. 

%%%%%%%%%%%%%%%%%%%%%%%%%%%%%%%%%%%%%%%%%%%%%%%%%%
\subsection{Perfectly inelastic collision ($e=0$ case)}

The vector $\vec{\varepsilon}$ presents a perfectly 
inelastic collision ($e=0$) between the two particles A and B. 
In this case, the two particles combine into a single 
particle which has mass $m_{A}+m_{B}$ and momentum 
$p_{A}+p_{B}$.

Since the masses are not altered after the collision in Newtonian 
mechanics, the tip of vector $\vec{\varepsilon}_{A}$ slides 
from point A to A${}^{\prime}$, as does 
$\vec{\varepsilon}_{B}$ from B to B${}^{\prime}$. 
Therefore, $\overrightarrow{\mathrm{OA}'}$ and 
$\overrightarrow{\mathrm{OB}'}$ 
indicate the vectors after collision and their addition 
$\overrightarrow{\mathrm{OA}'}+\overrightarrow{\mathrm{OB}'}$ 
becomes $\vec{\varepsilon}$. 
In figure \ref{fig:fig0203}, the lengths AA${}^{\prime}$ and 
BB${}^{\prime}$ 
describe the momentum lost by particle A and the momentum gained 
by particle B respectively, which are equal in length because of 
the conservation of the momentum. We obtain 
\begin{equation}
\textrm{AA}^{\prime} = \textrm{BB}^{\prime} 
= \dfrac{m_{A}m_{B}}{m_{A}+m_{B}}(v_{A}-v_{B}), 
\label{eq:lostgain}
\end{equation}
from (\ref{eq:amomentum}) with $e=0$ case.

%%%%%%%%%%%%%%%%%%%%%%%%%%%%%%%%%%%%%%%%%%%%%%%%%%
\subsection{Inelastic and elastic collision ($0 < e \leq 1$ case)}

The vector $\vec{\varepsilon}$ stays the 
same after collision in the case of $e \ne 0$, since the masses of 
the colliding particles do not change after collision 
and the momentum is conserved, that is, 
\begin{equation}
\vec{\varepsilon} = (m_{A}+m_{B}, p_{A}+p_{B}) 
= (m_{A}+m_{B}, p^{\prime\prime}_{A}+p^{\prime\prime}_{B}) 
= \vec{\varepsilon~}^{\prime\prime}. 
\end{equation}

\begin{figure}[htbp]
\centering
\begin{minipage}{0.4\columnwidth}
\centering
\includegraphics[width=\columnwidth]{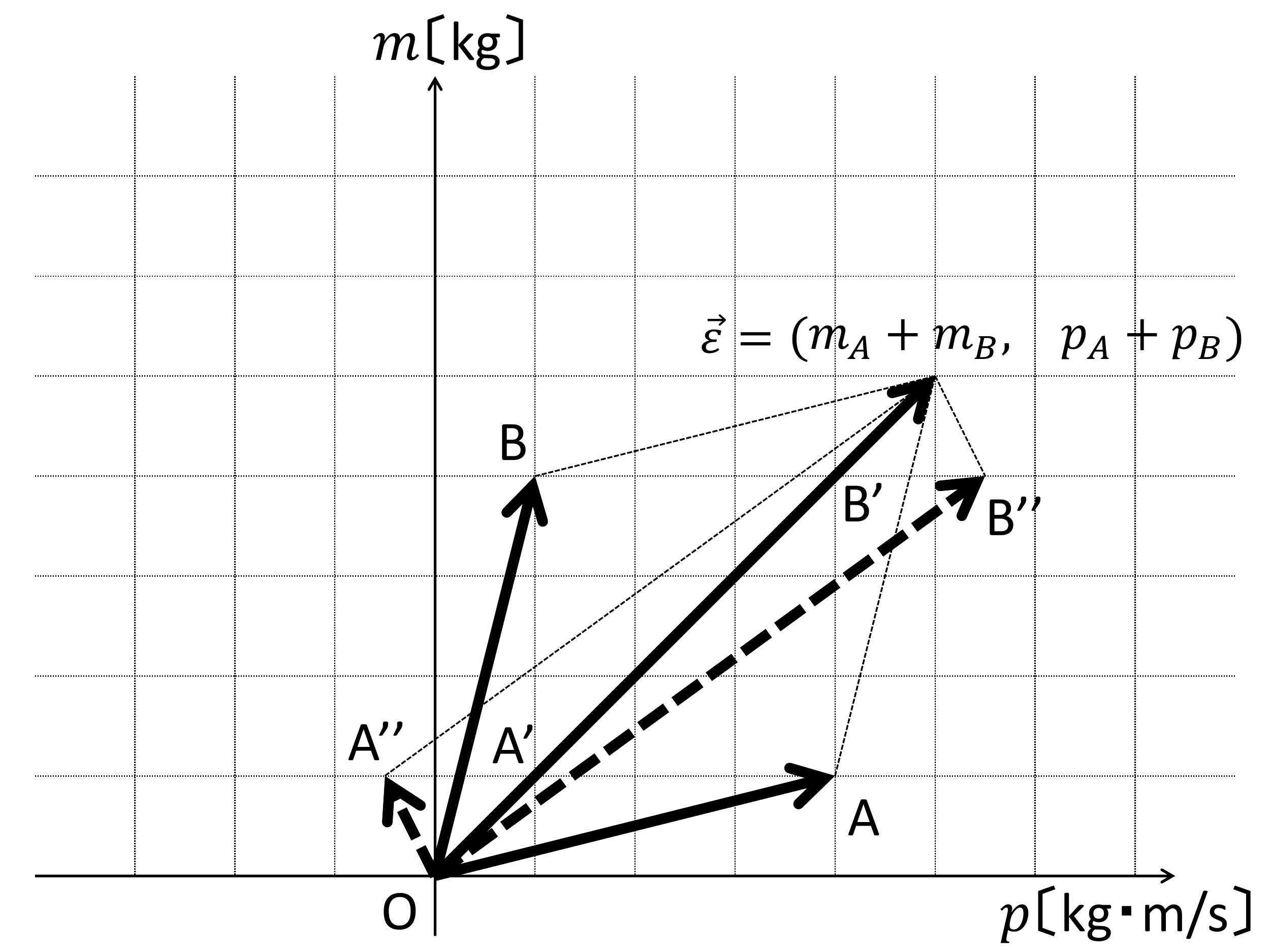}
\caption{The dashed vectors 
$\vec{\varepsilon~}^{\prime\prime}_{A}=(1, -0.5)$ and 
$\vec{\varepsilon~}^{\prime\prime}_{B}=(4, 5.5)$ are the 
mass-momentum vectors after collision with $e=0.5$}
\label{fig:fig04}
\end{minipage}
\hspace{5mm}
\begin{minipage}{0.4\columnwidth}
\centering
\includegraphics[width=\columnwidth]{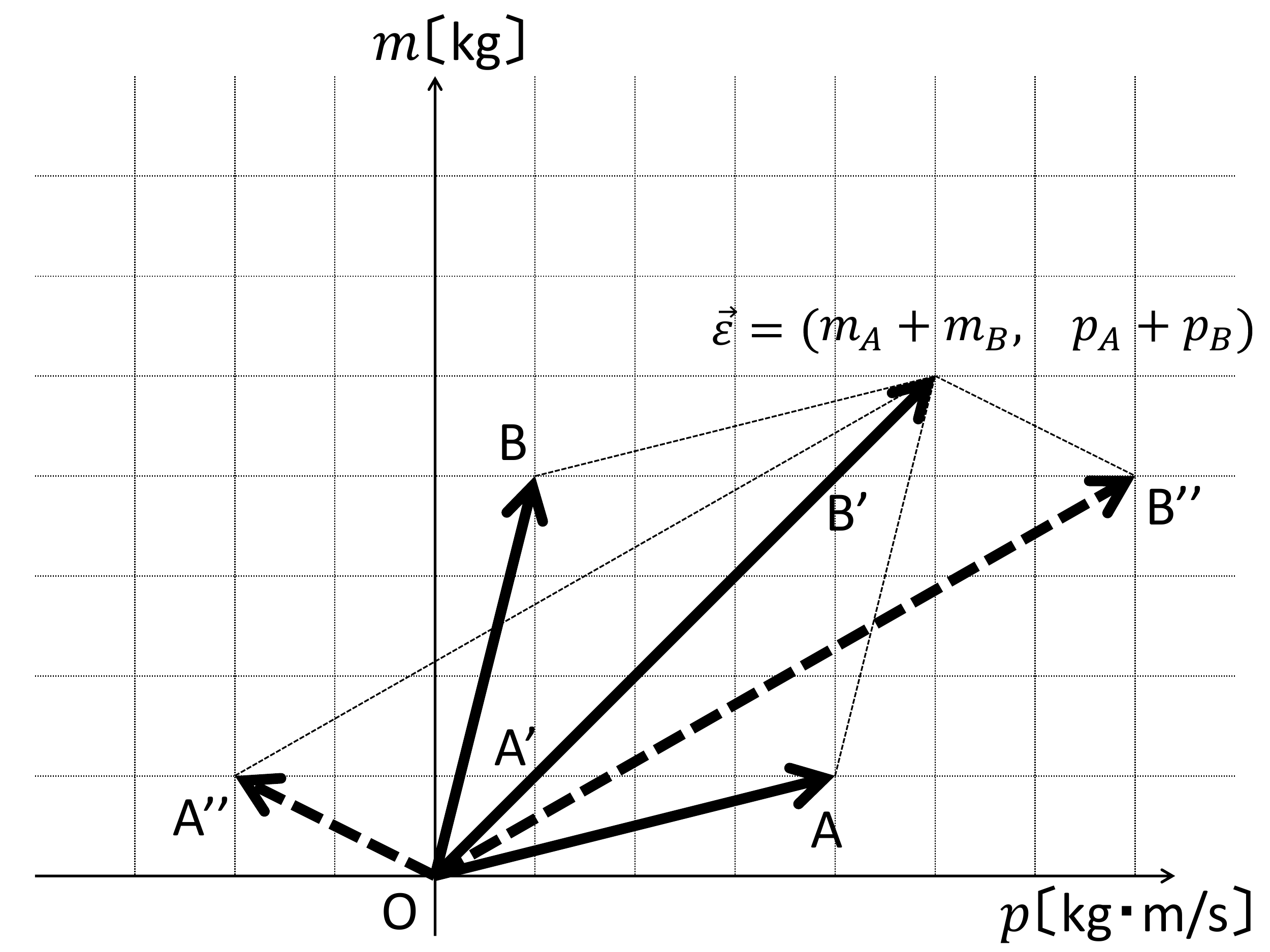}
\caption{The dashed vectors $\vec{\varepsilon~}^{\prime\prime}_{A}=(1, -2)$
 and $\vec{\varepsilon~}^{\prime\prime}_{B}=(4, 7)$ are the 
mass-momentum vectors after collision with $e=1$: elastic 
collision. }
\label{fig:fig05}
\end{minipage}
\end{figure}

We show the cases for $e = 0.5$ and $e = 1$ in figures \ref{fig:fig04} 
and \ref{fig:fig05}. The dashed vectors 
$\overrightarrow{OA''}=p^{\prime\prime}_{A}$ and 
$\overrightarrow{OB''}=p^{\prime\prime}_{B}$ show the mass-momentum 
vectors after collision with $e \ne 0$. 
We see from the figures that the addition of these dashed vectors 
gives the original vector $\vec{\varepsilon}$. 
In figures \ref{fig:fig04} and \ref{fig:fig05}, we can extract 
the following relationships between the lengths; 
\begin{equation}
\textrm{AA}^{\prime\prime} = \textrm{BB}^{\prime\prime}, 
\label{eq:eight}
\end{equation}
\begin{equation}
\textrm{A}^{\prime}\textrm{A}^{\prime\prime} 
= e \times \textrm{AA}^{\prime}, \qquad 
\textrm{B}^{\prime}\textrm{B}^{\prime\prime} 
= e \times \textrm{BB}^{\prime}. 
\label{eq:tips}
\end{equation}
Equation (\ref{eq:eight}) signifies the momentum conservation. 
AA${}^{\prime\prime}$ is the momentum lost of particle A and 
BB${}^{\prime\prime}$ is the momentum gain of particle B. 
Equation (\ref{eq:tips}) are understood by the second term 
in the right hand side of (\ref{eq:amomentum}). 
A first part of it shows 
$\textrm{AA}^{\prime}=\textrm{BB}^{\prime}=
\tfrac{m_{A}m_{B}}{m_{A}+m_{B}}(v_{A}-v_{B})$, which is already 
described by (\ref{eq:lostgain}). In addition to this, another part 
$\textrm{A}^{\prime}\textrm{A}^{\prime\prime}
=\textrm{B}^{\prime}\textrm{B}^{\prime\prime}
=e\tfrac{m_{A}m_{B}}{m_{A}+m_{B}}(v_{A}-v_{B})$ signifies 
$\textrm{AA}^{\prime}=\textrm{BB}^{\prime}$ times the coefficient of 
restitution $e$, which takes the value from 0 to 1. So the length 
$\textrm{A}^{\prime}\textrm{A}^{\prime\prime}
=\textrm{B}^{\prime}\textrm{B}^{\prime\prime}$ are changed according 
to the value of $e$. When the collision problem we concern gives 
the value of $e$, we can immediately fix the points 
$\textrm{A}^{\prime\prime}$ and $\textrm{B}^{\prime\prime}$ 
according to (\ref{eq:tips}).

%%%%%%%%%%%%%%%%%%%%%%%%%%%%%%
%%%%%%%%%%%%%%%%%%%%%%%%%%%%%%

Figure \ref{fig:fig05} presents the complete picture of a one 
dimensional collision 
without any calculation. During the collision, the vectors 
$\vec{\varepsilon}_{A}$ and $\vec{\varepsilon}_{B}$ 
slide in parallel to the 
horizontal line because of the invariance of the masses. 
After the collision, the tips of the vectors lie between 
A${}^{\prime}$A${}^{\prime\prime}$ and 
B${}^{\prime}$B${}^{\prime\prime}$ according to (\ref{eq:tips}). 
Once we fix the value of the coefficient of 
restitution $e$, we write down the two vectors 
$\vec{\varepsilon~}^{\prime\prime}_{A}$ and 
$\vec{\varepsilon~}^{\prime\prime}_{B}$ 
immediately. The addition of these vectors gives 
the original vector $\vec{\varepsilon}$. 

As a homework or exam problems to these diagrams, let us 
consider a moving multistage rocket which gets rid of a stage. 
The remaining rocket goes to the same direction, while the stage 
goes to the opposite direction. 
This problem is depicted like figure \ref{fig:fig05}. 
The original rocket is assigned to $\vec{\varepsilon}$, the stage 
is for A${}^{\prime\prime}$ and the remaining rocket is for 
B${}^{\prime\prime}$. This disintgration can be viewed the 
time-reverse process of an $e=0$ collision. 

%%%%%%%%%%%%%%%%%%%%%%%%%%%%%%%%%%%%%%%%
\subsection{Center-of-mass systems}

In collision problems, we often see the collisions from the 
point of view of the center-of-mass system. Since the sum of the 
momentum is zero before and after collision, the equations become 
simpler, compared to the laboratory system. 

\begin{figure}[htbp]
\centering
\begin{minipage}{0.4\columnwidth}
\centering
\includegraphics[width=\columnwidth]{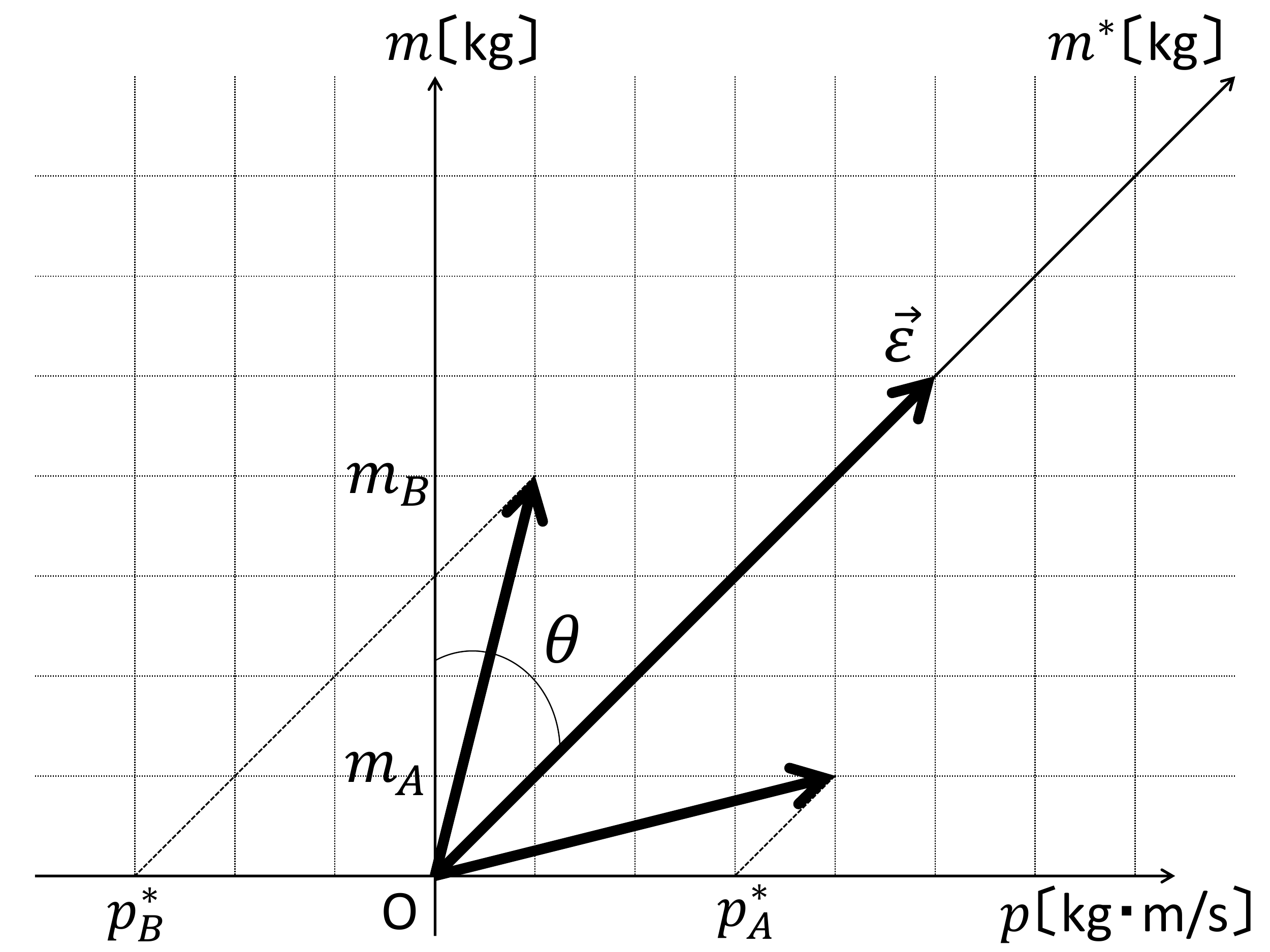}
\caption{We draw a line from the tips of the vectors to the 
horizontal axis parallel to the vector $\vec{\varepsilon}$. 
The crossing points are the momenta in the center of mass system. 
Eqs. (\ref{eq:bmomc}) show $p_{A}^{\ast} = -p_{B}^{\ast} = 3$. }
\label{fig:fig06}
\end{minipage}
\hspace{5mm}
\begin{minipage}{0.4\columnwidth}
\centering
\includegraphics[width=\columnwidth]{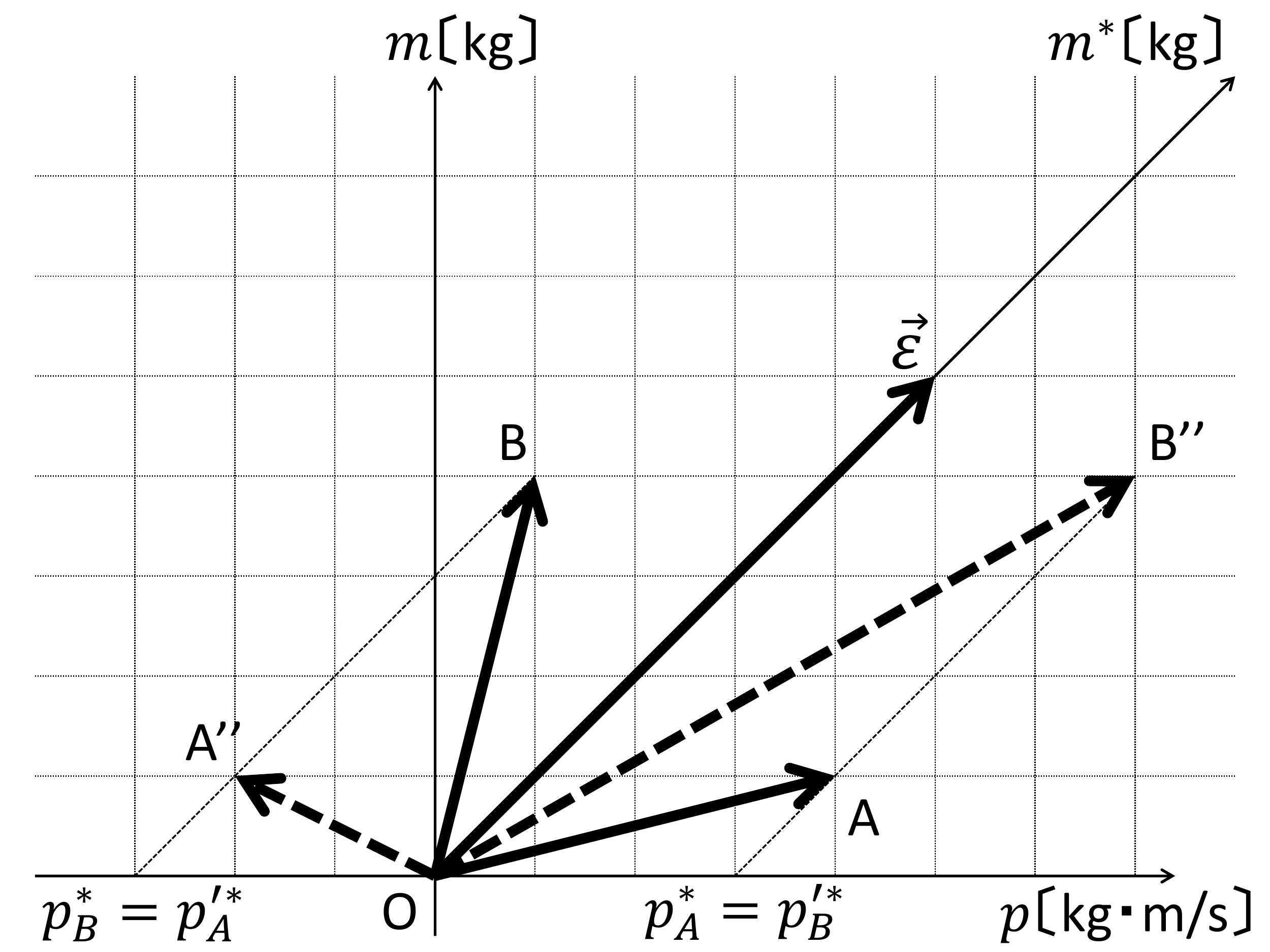}
\caption{The dashed vectors are the mass-momentum vectors 
after collision with $e=1$: elastic collision case. Eqs. 
(\ref{eq:amomc}) show $-p^{\prime\ast}_{A} = p^{\prime\ast}_{B} = 3$. }
\label{fig:fig07}
\end{minipage}
\end{figure}

The speed of the center-of-mass system $V$ is given by 
\begin{equation}
V = \dfrac{p_{A}+p_{B}}{m_{A}+m_{B}} 
= \dfrac{m_{A}v_{A}+m_{B}v_{B}}{m_{A}+m_{B}}=\tan\theta, 
\end{equation}
where $\theta$ is depicted in figure \ref{fig:fig06} and describes 
the slope of 
the vector $\vec{\varepsilon}$ relative to the vertical axis $m$. 

%%%%%%%%%%%%%%%%%%%%%%%%%%%%%%
%%%%%%%%%%%%%%%%%%%%%%%%%%%%%%

We assign an $\ast$ to the variables which describe collisions in 
the center-of-mass system. The velocities before collision 
are obtained 
\begin{equation}
\begin{cases}
v^{\ast}_{A} &= v_{A}-V = +\dfrac{m_{B}}{m_{A}+m_{B}}(v_{A}-v_{B}), \\
v^{\ast}_{B} &= v_{B}-V = -\dfrac{m_{A}}{m_{A}+m_{B}}(v_{A}-v_{B}), 
\end{cases}
\end{equation}
and the momenta are 
\begin{equation}
\begin{cases}
p^{\ast}_{A} &= m_{A}v^{\ast}_{A} 
              = +\dfrac{m_{A}m_{B}}{m_{A}+m_{B}}(v_{A}-v_{B}), \\
p^{\ast}_{B} &= m_{B}v^{\ast}_{B} 
              = -\dfrac{m_{A}m_{B}}{m_{A}+m_{B}}(v_{A}-v_{B}). 
\end{cases}
\label{eq:bmomc}
\end{equation}
From (\ref{eq:avelo}), we obtain the velocities and momenta 
after collision as 
\begin{equation}
\begin{cases}
v^{\prime\ast}_{A} 
&= v^{\prime}_{A}-V = -e\dfrac{m_{B}}{m_{A}+m_{B}}(v_{A}-v_{B}), \\
v^{\prime\ast}_{B} 
&= v^{\prime}_{B}-V = +e\dfrac{m_{A}}{m_{A}+m_{B}}(v_{A}-v_{B}), 
\end{cases}
\end{equation}
and 
\begin{equation}
\begin{cases}
p^{\prime\ast}_{A} &= -e\dfrac{m_{A}m_{B}}{m_{A}+m_{B}}(v_{A}-v_{B}), \\
p^{\prime\ast}_{B} &= +e\dfrac{m_{A}m_{B}}{m_{A}+m_{B}}(v_{A}-v_{B}). 
\end{cases}
\label{eq:amomc}
\end{equation}
Below we show how we obtain these results from the mass-momentum 
diagram without requiring any calculations. Figure \ref{fig:fig06} 
shows the diagram 
before collision, which is the same diagram as figure \ref{fig:fig0203}. 
We draw straight lines from the tips of the vectors 
$\vec{\varepsilon}_{A}$ and $\vec{\varepsilon}_{B}$ to the horizontal 
axis $p$, parallel to the 
vector $\vec{\varepsilon}$. The points of intersection show the 
momenta in the center-of-mass system, $p^{\ast}_{A}$ and 
$p^{\ast}_{B}$. Moreover, $p^{\ast}_{A}$ and $p^{\ast}_{B}$ are always 
positioned symmetrically about the origin. So, 
$p^{\ast}_{A}+p^{\ast}_{B}=0$ is understood at a glance. 
Figure \ref{fig:fig07} shows the elastic collision $e=1$ case, 
which is the same diagram as in figure \ref{fig:fig05}. 
For elastic collisions, the momenta of the particles are exchanged 
after the collisions. As we see from figure \ref{fig:fig07}, 
$p^{\ast}_{A}=p^{\prime\ast}_{B}$, $p^{\ast}_{B}=p^{\prime\ast}_{A}$ 
and 
$p^{\ast}_{A}+p^{\ast}_{B}=p^{\prime\ast}_{A}+p^{\prime\ast}_{B}=0$ 
are fulfilled. 

The direction of $\vec{\varepsilon}$ 
in figure \ref{fig:fig07} constructs an oblique coordinate system 
with the horizontal axis $p$.
This oblique coordinate system describes the state of the particles in 
the center-of-mass system. 
Oblique coordinates are often used in special relativity to describe 
events from different reference frames~\cite{takeuchi}.

%%%%%%%%%%%%%%%%%%%%%%%%%%%%%%%%%%%%%%%%
\section{Laboratory systems}

\begin{figure}[htbp]
\centering
\includegraphics[width=7.5cm]{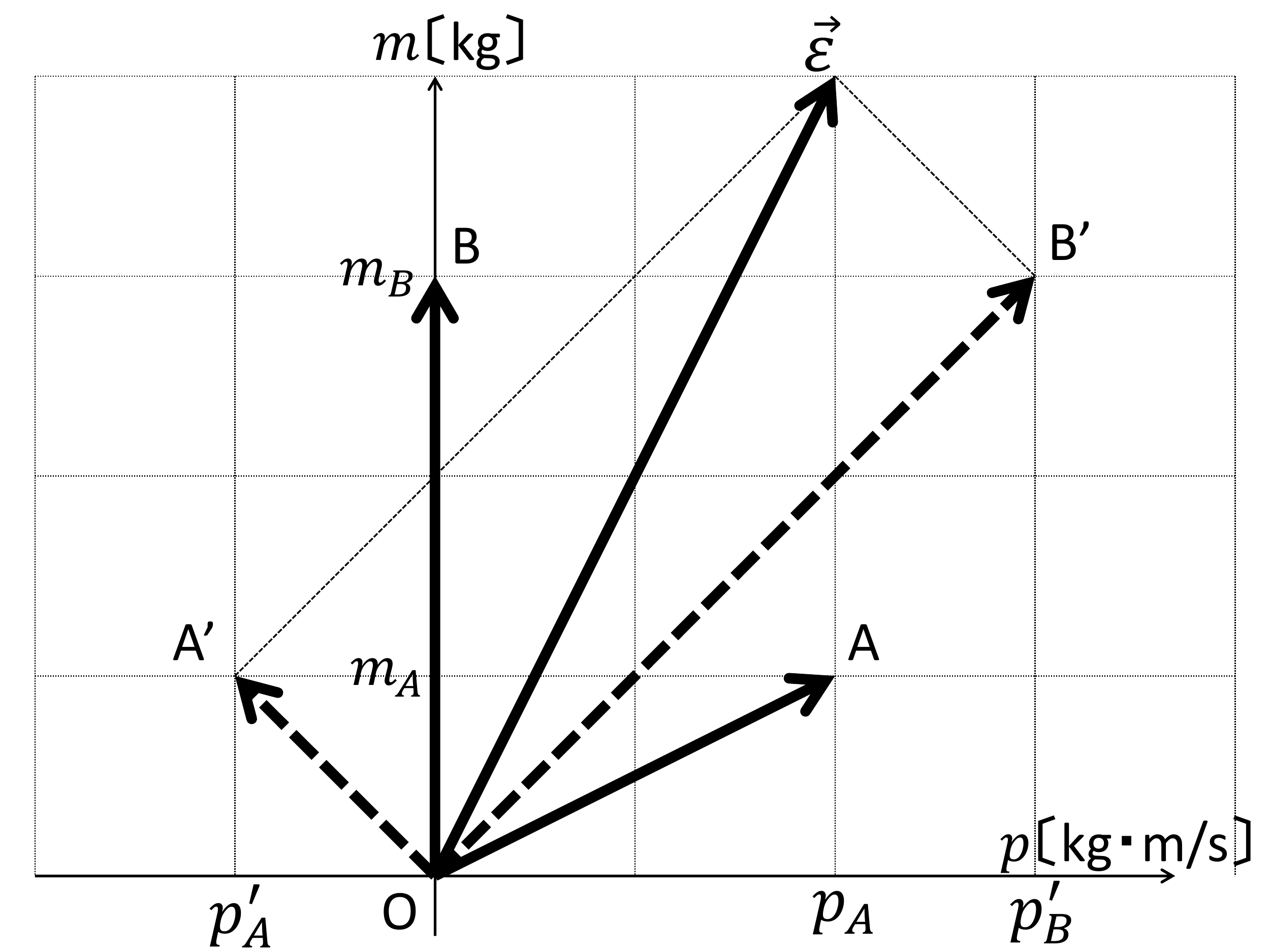}
\caption{Two vectors in the laboratory system. The case 
$\vec{\varepsilon}_{A}=(m_{A}, p_{A})=(1, 2)$ and 
$\vec{\varepsilon}_{B}=(m_{B}, p_{B})=(3, 0)$ means $v_{A}=2$ and 
$v_{B}=0$. The addition of $\vec{\varepsilon}_{A}$ and 
$\vec{\varepsilon}_{B}$ describes a perfectly inelastic collision. 
The dashed vectors are the mass-momentum vectors after collision 
with $e=1$: elastic collision case.}
\label{fig:fig0809}
\end{figure}

In this section, we utilize the foregoing methods for collisions in 
laboratory systems. We consider the system in which the particle B 
is at rest before the collision. This case is depicted in 
figure \ref{fig:fig0809}. The vector $\vec{\varepsilon}_{B}$ is along 
the vertical axis. The vector $\vec{\varepsilon}$ is the addition of 
$\vec{\varepsilon}_{A}$ and $\vec{\varepsilon}_{B}$ and describes 
the state of the particle in the case of a perfectly inelastic collision 
$(e=0)$. 

%%%%%%%%%%%%%%%%%%%%%%%%%%%%%%
%%%%%%%%%%%%%%%%%%%%%%%%%%%%%%

We also show the elastic collision ($e=1$) case in figure 
\ref{fig:fig0809}. The dashed vectors are the states of the colliding 
particles after collision and the adding them gives 
$\vec{\varepsilon}$. 
Moreover, AA${}^{\prime}$=BB${}^{\prime}$ due to the 
conservation of momentum.

\begin{figure}[htbp]
\centering
\includegraphics[width=7.5cm]{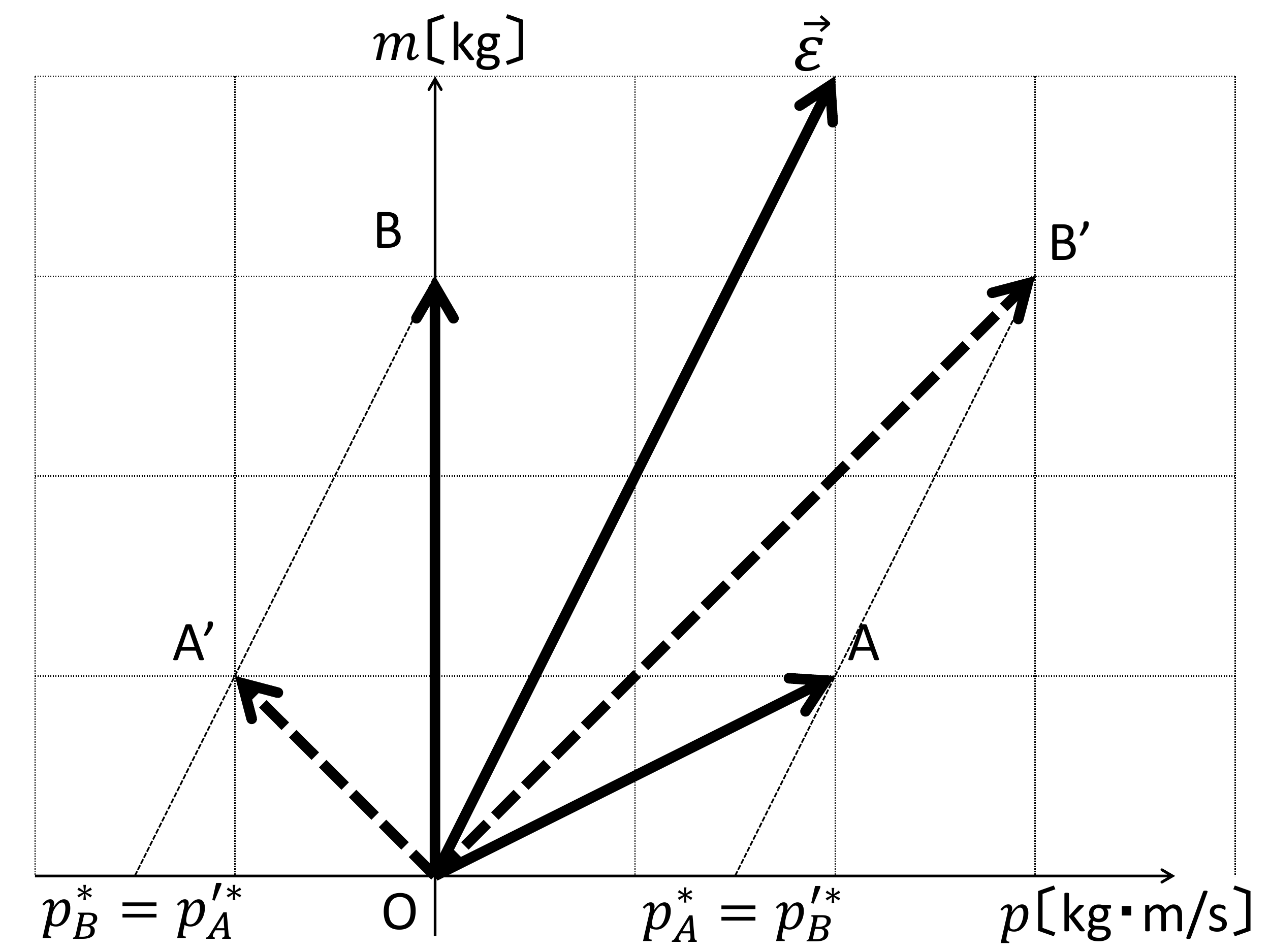}
\caption{We draw a line from the tips of the vectors to the 
horizontal axis parallel to $\vec{\varepsilon}$. The 
points at which these cross the horizontal axis 
give the momenta in the center-of-mass system. 
Eqs. (\ref{eq:bmomc}) and (\ref{eq:amomc}) show 
$p_{A}^{\ast} = -p_{B}^{\ast} = 1.5 =- p_{A}^{\prime\ast} 
= p_{B}^{\prime\ast}$. }
\label{fig:fig10}
\end{figure}

We see this collision from the center-of-mass system depicted in 
figure \ref{fig:fig10}. This figure is the same as 
figure \ref{fig:fig0809}. 
For the elastic collision ($e=1$), the momenta of 
the colliding particles are exchanged after collision. 
This can be seen in the figure at a glance. 

%%%%%%%%%%%%%%%%%%%%%%%%%%%%%%
%%%%%%%%%%%%%%%%%%%%%%%%%%%%%%

%%%%%%%%%%%%%%%%%%%%%%%%%%%%%%%%%%%%%%%%%%%%%%%%%%%%%%%%%%%%%%%%%%%%%%
\section{Summary}

We introduce the mass-momentum diagram to analyze collision 
problems in Newtonian mechanics. This diagram serves as a powerful 
tool for obtaining the momenta of particles after a collision without 
the need for any calculations. 
The same diagram can also be used to analyze collisions from the 
point of view of the center-of-mass system. Students are often 
confusing in transposition between the laboratory system and the 
center-of-mass system. 
A single diagram can give students not only intuitive point of view, 
but also qualitative description of all features of the collisions 
without any calculations. 

As a homework or exam problem, let us analyze a disintegration of 
a particle into two parts which move independently to opposite 
direction after the disintegration. The orignal particle is drawn 
along the vertical axis in which the particle is at rest before 
the disintegration. 
After the disintegration, the two parts are drawn symmetrically 
about the vertical axis. Even if the masses of the two parts are 
different, the addition of those vectors become the same as 
the original particle before the disintegration. 
This example is the same case with the rocket which is mentioned 
at the end of subsection 3.2. However, we should draw a different 
diagram according to the reference frame which we observe the parent 
body. 

Finally, the use of these diagrams can be extended to two dimensional 
case in a future paper. 

%%%%%%%%%%%%%%%%%%%%%%%%%%%%%%%%%%%%%%%%%%%%%%%%%%%%%%%%%%%%%%%%%%%%%%
\section*{Acknowledgemnt}

The author thanks the anonymous reviewers for their helpful suggestions. 

%%%%%%%%%%%%%%%%%%%%%%%%%%%%%%%%%%%%%%%%%%%%%%%%%%%
\vspace{1cm}

%%%%%%%%%%%%%%%%%%%%%%%%%%%%%%%%%%%%%%%%%%%%%%%%%%%
%%%%%%%%%%%%%%%%%%%%%%%%%%%%%%%%%%%%%%%%%%%%%%%%%%%
%%%%%%%%%%  REFERENCES  %%%%%%%%%%%%%%%%%%%%%%%%%%%
%%%%%%%%%%%%%%%%%%%%%%%%%%%%%%%%%%%%%%%%%%%%%%%%%%%
%%%%%%%%%%%%%%%%%%%%%%%%%%%%%%%%%%%%%%%%%%%%%%%%%%%

%%%%%%%%%%%%%%%%%%%%%%%%%%%%%%%%%%%%%%%%%%%%%%%%%%%
%%%%%%%%%%%%%%%%%%%%%%%%%%%%%%%%%%%%%%%%%%%%%%%%%%%

\begin{thebibliography}{123}
%%%%%%%%%%%%%%%%%%%%%%%%%%%%
\bibitem{saletan} 
Saletan E. J. 1997 Minkowski diagram in momentum space, 
\textit{Am. J. Phys.} {\bf 65} 799-800.

\bibitem{bokor} 
Bokor N. 2011 Analysing collisions using Minkowski diagram in 
momentum space, 
\textit{Eur. J. Phys.} {\bf 32} 773-782.

\bibitem{landau2} 
Landau L. D. and Lifshitz E. M. 1976 
\textit{The Classical Theory of Fields}, 
(Butterworth-Heinenann).

\bibitem{takeuchi} 
Takeuchi T. 2010 
\textit{An Illustrated Guide to Relativity}, 
(Cambridge).

%%%%%%%%%%%%%%%%%%%%%%%%%%%%
\end{thebibliography}
\end{document}